\documentclass[
    ,final            % use final for the camera ready runs
  ]
  {aipproc}

\layoutstyle{6x9}

\newcommand{\figheight}{0.86\hsize}
\newcommand{\etal}{{\it et al.}}

\begin{document}

\title{Chiral behavior of baryon magnetic moments}


\author{Derek B. Leinweber}{
  address={Special Research Centre for the Subatomic Structure of Matter
           and School of Chemistry and Physics, University of Adelaide, 
           Adelaide SA 5005, Australia}
}

\author{Ross D. Young}{
  address={Special Research Centre for the Subatomic Structure of Matter
           and School of Chemistry and Physics, University of Adelaide, 
           Adelaide SA 5005, Australia}
}

\begin{abstract}
   The utility of chiral effective field theory, constructed in a
   manner in which loop contributions are suppressed as one moves
   outside the power-counting regime, is explored for baryon magnetic
   moments.  Opportunities for the study of significant chiral
   curvature in valence and full QCD and the nontrivial behavior of
   strange- and light-quark contributions to the magnetic moment of
   the $\Lambda$ baryon are highlighted.
\end{abstract}

\maketitle

%%%%%%%%%%%%%%%%%%%%%%%%%%%%%%%%%%%%%%%%%%%%
%% MAINMATTER
%%%%%%%%%%%%%%%%%%%%%%%%%%%%%%%%%%%%%%%%%%%%

\section{Introduction}

It is a pleasure to have this opportunity to review some of the most
interesting aspects of baryon structure on the occasion of Tony
Thomas' 60th birthday.  In reflecting on our collaborative work with Tony [1--70]
% \cite{Leinweber:1999ig,%
% Leinweber:1998ej,%
% Leinweber:2003dg,%
% Young:2002cj,%
% Leinweber:2004tc,%
% Young:2002ib,%
% Leinweber:2001ac,%
% Leinweber:2001ui,%
% Leinweber:1999nf,%
% HackettJones:2000qk,%
% HackettJones:2000js,%
% Leinweber:2006ug,%
% Young:2006jc,%
% Cloet:2003jm,%
% Young:2004tb,%
% Ashley:2003sn,%
% Flambaum:2004tm,%
% Leinweber:2000sa,%
% Lasscock:2005tt,%
% Allton:2005fb,%
% Detmold:2001hq,%
% Wang:2007iw,%
% Young:2007zs,%
% Thomas:2004iw,%
% Leinweber:2002bw,%
% Lasscock:2005kx,%
% Cloet:2002eg,%
% Leinweber:2005xz,%
% Young:2009zb,%
% Giedt:2009mr,%
% Leinweber:2003ux,%
% Young:2003ns,%
% Leinweber:2005bz,%
% %Young:2001nc,%
% Young:2003gd,%
% Wang:2005vg,%
% Armour:2005mk,%
% Wang:2004gz,%
% Wang:1900ta,%
% Thomas:2005qm,%
% Wright:2000gg,%
% Young:2002rx,%
% Leinweber:1999bv,%
% Thomas:1999ae,%
% %Cloet:2002sx,%
% Wang:2004wja,%
% Wang:2004vg,%
% Wang:2008vb,%
% Thomas:1999mv,%
% Leinweber:2000yf,%
% Wright:2001ip,%
% Thomas:2003pd,%
% Thomas:2005qb,%
% Carroll:2008sv,%
% Leinweber:2000pw,%
% Morel:2003fj,%
% Thomas:2006jf,%
% Wright:2002vn,%
% %Young:2002sw,%
% Leinweber:2003zy,%
% Thomas:2003te,%
% %Wang:2004gya,%
% Young:2005tr,%
% Leinweber:2005cm,%
% %Thomas:2005ah,%
% Allton:2005vm,%
% Leinweber:2005zj,%
% Lasscock:2004kp,%
% Armour:2008ke,%
% Young:2009ps,%
% Thomas:2003ga,%
% Leinweber:2005jb,%
% Wang:2005tt,%
% %Young:2003ga,%
% Lasscock:2005wb,%
% Thomas:2007zzd}
%
it is clear that the focus must be on the development of finite-range
regularised (FRR) chiral effective field theory ($\chi$EFT); a
breakthrough founded on deep insights gained through an extensive
foundation of modeling hadron phenomenology now vindicated through an
analysis of the most recent lattice QCD results \cite{Hall:2010ai}.

In this presentation, we draw on the lattice simulation results of the
CSSM Lattice Collaboration for baryon electromagnetic form factors
published in Ref.~\cite{Boinepalli:2006xd}.  These results approaching
the quenched chiral limit were made possible through the improved
chiral properties of the FLIC fermion action
\cite{Leinweber:2002bw,Zanotti:2001yb,Zanotti:2004dr,Boinepalli:2004fz}.

\section{The P\'ade}

Through the application of the Cloudy Bag Model to the
extrapolation of lattice QCD results \cite{Leinweber:1999ig}, it became
abundantly clear that nearly all of the lattice QCD results of the
time were outside of the power counting regime of chiral perturbation
theory \cite{Leinweber:2005cm}.  This lead to the development of
chiral extrapolation techniques that contain the correct non-analytic
behavior of chiral perturbation theory but take on the smooth almost
linear behavior of baryon properties at moderately large quark
masses. 
The first approach exploited a P\'ade \cite{Leinweber:1998ej}
\begin{equation}
\mu_{p(n)}=\frac{\mu _{0}}{1 - \chi_{p(n)} \, m_{\pi } / {\mu_{0}} + \beta
\, m^{2}_{\pi }} \, ,
\label{mag-mom}
\end{equation}
which builds in the Dirac moment at moderately large $m^{2}_{\pi }$
and has the correct leading non-analytic (LNA) behavior of chiral
perturbation theory at small $m_\pi$, $\mu =\mu _{0} + \chi_{p(n)} m_{\pi} +
\cdots$, with $\chi_{p(n)}$ a known model independent constant.  While this
approach was phenomenologically successful, it became clear that
incorporating higher order terms and accommodating baryons having the
coefficient $\chi > 0$ was difficult
\cite{Leinweber:1999nf,HackettJones:2000qk}.

\section{Finite Range Regularised Effective Field Theory}

The solution was found in an a alternative regularisation of chiral
effective field theory now known as Finite Range Regularisation where
a regulator function is introduced to suppress the large momenta of
effective degrees of freedom in chiral loop integrals
\cite{Leinweber:2003dg,Young:2002ib,Young:2004tb,Young:2003ns}.
Mathematically equivalent to chiral perturbation theory to any finite
order calculated, this series expansion has the additional feature of
the suppression of loop-integral contributions at moderate pion masses
in accord with lattice simulation results.  Thus, this resummation of
the chiral expansion will not make catastrophic errors outside of the
power-counting regime.

Turning our focus now to the specific phenomenology of baryon magnetic
moments, we employ the approach of Ref.~\cite{Leinweber:2004tc} which
includes the non-analytic behavior from photon couplings to $\pi$ and
$K$ mesons in the presence of intermediate octet or decuplet baryons.
Unphysical $\eta'$ contributions are also included in the quenched
analysis.  Whereas Ref.~\cite{Leinweber:2004tc} focused on the
quark-sector contributions to baryon magnetic moments and their
environment sensitivity, here we combine the sectors to study the
chiral behavior of the proton and neutron magnetic moments.

As the lattice simulation results \cite{Boinepalli:2006xd} are
obtained in the quenched approximation, we exploit our discovery of a
link between quenched QCD and full QCD through a replacement of the
quenched meson cloud with the meson cloud of full QCD
\cite{Young:2002cj} via FRR $\chi$EFT \cite{Leinweber:2003dg}.  The
coefficients for unquenching baryon magnetic moments are taken from
Ref.~\cite{Leinweber:2002qb}.

\begin{figure}[tbp]
{\includegraphics[height=\figheight,angle=90]{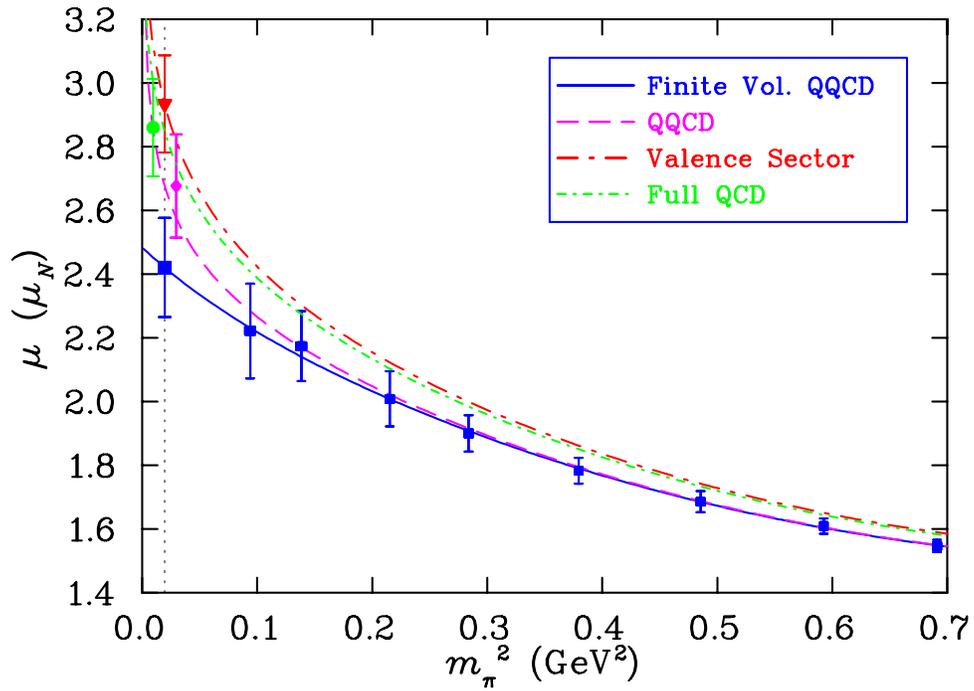}}
\vspace*{-0.5cm}
\caption{
The magnetic moment of the proton.  Curves and symbols are described
in the text.
}
\label{proton}
\vspace{-0.5cm}
\end{figure}

\begin{figure}[tbp]
{\includegraphics[height=\figheight,angle=90]{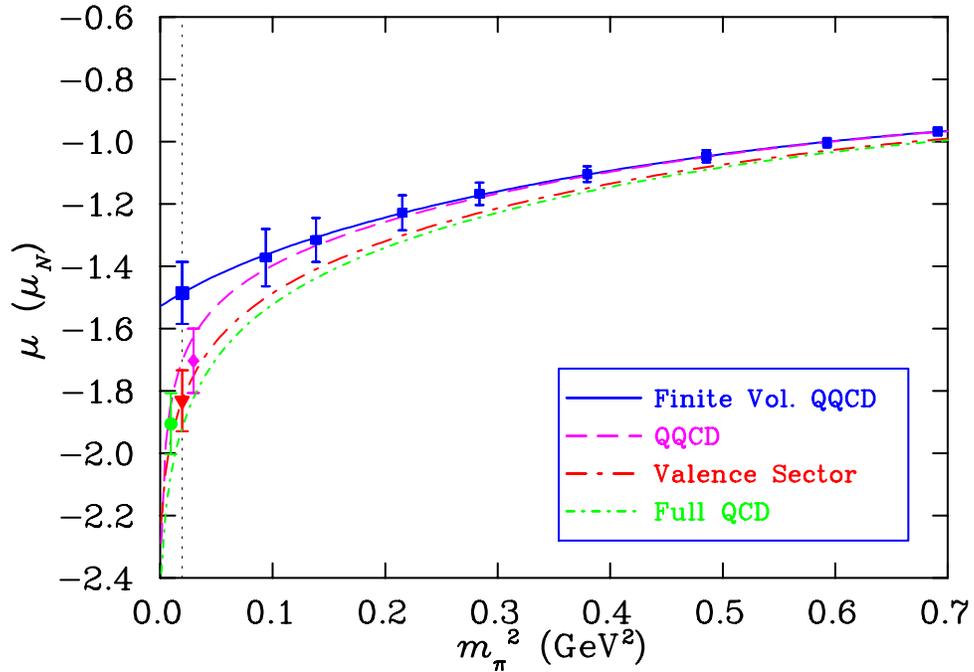}}
\vspace*{-0.5cm}
\caption{
The magnetic moment of the neutron.  Curves and symbols are described
in the text.
}
\label{neutron}
\vspace{-0.5cm}
\end{figure}

Figures \ref{proton} and \ref{neutron} illustrate results for the
proton and neutron respectively.
A fit of FRR quenched $\chi$EFT (solid curve) to the FLIC fermion
lattice points is illustrated.  Here only the discrete momenta allowed
in the finite volume of the lattice are summed in performing the loop
integral.  The long-dashed curve that also runs through the lattice
results corresponds to replacing the discrete momentum sum by the
infinite-volume, continuous momentum integral.  Incorporating baryon
mass splittings into the kaon loop contributions is essential -- e.g.,
the contribution of $\Sigma \to N K$ is almost doubled when the
$\Sigma - N$ mass splitting is included.

The removal of quenched $\eta'$ contributions and the appropriate
adjustment of $\pi$ and $K$ loop coefficients~\cite{Leinweber:2002qb}
provides the dot-dash curve of Fig.~\ref{proton}.  This is our best
estimate of the proton magnetic moment of full QCD where only the
valence quarks couple to the electromagnetic current; i.e. a connected
insertion of the electromagnetic current.  This valence contribution
can be thought of as a form of partially-quenched QCD where the sea
quarks have the correct mass, but have neutral electric charges.

Finally, contributions from the disconnected insertion of the current
are estimated via meson loop contributions to provide the proton
magnetic moment in full QCD (fine dash-dot curve in
Fig.~\ref{proton}).  Figure~\ref{neutron} displays similar results for
the neutron.

The calculation of valence properties in full QCD is a relatively easy
task as the disconnected current insertions are not required.
Moreover the chiral non-analytic behavior in valence QCD is strong as
the screening effects from anti-quark couplings in the mesons are
absent.  Noting that disconnected current insertions vanish in
isovector contributions, valence QCD and full QCD are equivalent for
isovector quantities, making these observables ideal for revealing and
understanding the meson cloud of baryons.  We predict significant
enhancements in the magnitude of proton and neutron magnetic moments
on large-volume lattices as the chiral limit is approached.

\begin{figure}[tbp]
{\includegraphics[height=\figheight,angle=90]{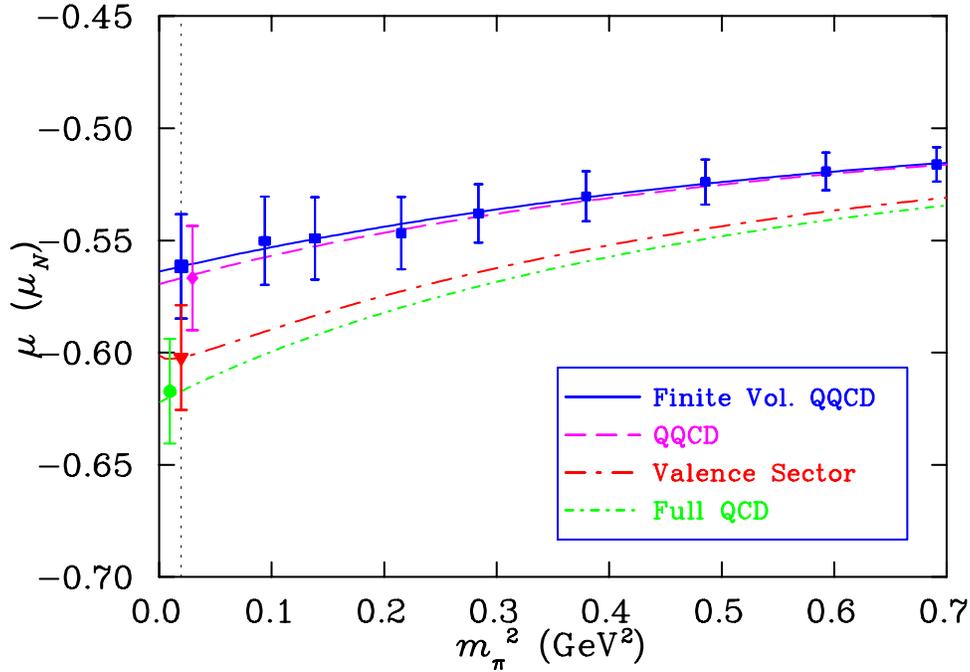}}
\vspace*{-0.5cm}
\caption{
The magnetic moment of the $\Lambda$.  Curves and symbols are described
in the text.
\vspace{-0.5cm}
}
\label{Lambda}
\end{figure}

The magnetic moment of the $\Lambda$ baryon, illustrated in
Fig.~\ref{Lambda}, is particularly interesting to study.  In simple
quark models the moment is provided solely by the strange constituent
quark.  However, these recent precise lattice results
\cite{Boinepalli:2006xd} reveal it to have a more interesting
composition.  First, we note how the magnetic moment has a clear
dependence on the quark mass, represented by $m_\pi^2$.  Whereas $\pi
\Sigma$ contributions survive in valence QCD, they vanish in quenched
and full QCD, leaving kaon contributions to drive the non-analytic
behavior.  While there is little curvature from the kaons, we note
that the unquenching of the kaon contribution from quenched to full
QCD is significant.

\begin{figure}[tbp]
{\includegraphics[height=\figheight,angle=90]{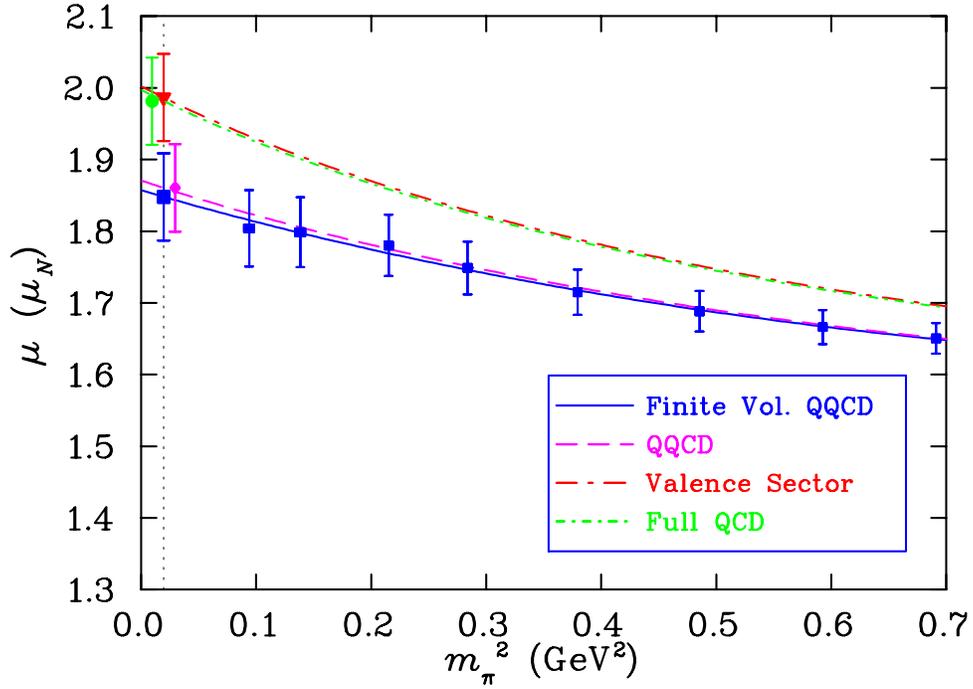}}
\vspace*{-0.5cm}
\caption{
Contribution of the strange quark to the magnetic moment of the
$\Lambda$ normalized to unit charge.
\vspace{-0.5cm}
}
\label{sInLambda}
\end{figure}

Figure \ref{sInLambda} illustrating the strange-quark contribution to
the $\Lambda$ magnetic moment reveals the $m_\pi^2$ dependence to be
predominantly due to an environmental dependence of the strange quark
contribution.  While the mass of the strange quark is held fixed in
the simulations, the changing light quark mass leads to a change in
the strange quark contribution.  Such environmental effects are natural
in terms of $K N$ and $K \Xi$ dressings of the Lambda and unquenching
effects are predicted to be significant.

\begin{figure}[tbp]
{\includegraphics[height=\figheight,angle=90]{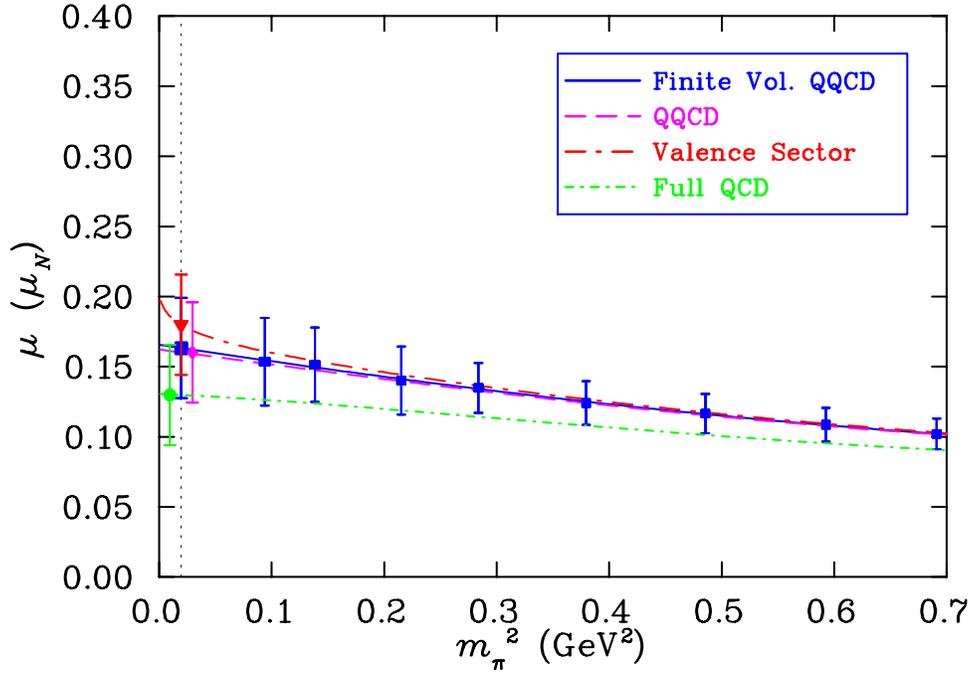}}
\vspace*{-0.5cm}
\caption{
Contribution of a $u$ or $d$ quark to the magnetic moment of the
$\Lambda$ normalized to unit charge.
\vspace{-0.5cm}
}
\label{uInLambda}
\end{figure}

Figure \ref{uInLambda} reveals the nontrivial role of light-quark
contributions to the Lambda magnetic moment.  Again, $\pi \Sigma$
contributions are admitted in valence QCD \cite{Leinweber:2002qb},
whereas kaon dressings drive the non-analytic
behavior of quenched and full QCD.

In summary, FRR $\chi$EFT has provided many precise predictions for
the electromagnetic properties of baryons in both valence and full
QCD.  With large-volume lattice QCD simulations of valence QCD now at
the fore of the field, it will be interesting to learn the success of
these predictions and the extent to which our current understanding of
the essential physics behind these observables is accurate.

\begin{theacknowledgments}
We thank Wally Melnitchouk for his leadership in organising this
festive workshop.  Aspects of this research were undertaken on the NCI
National Facility in Canberra, Australia, which is supported by the
Australian Commonwealth Government.  We also thank eResearch SA for
generous grants of supercomputer time which have enabled this project.
This work is supported by the Australian Research Council.
\end{theacknowledgments}

%%%%%%%%%%%%%%%%%%%%%%%%%%%%%%%%%%%%%%%%%%%

\bibliographystyle{aipproc}   % if natbib is available

\end{document}